\documentclass[12pt,preprint]{aastex}
\usepackage{color}
\shortauthors{Matthews et al.}
\shorttitle{The Radio Photosphere of Mira}

\begin{document}
\newcommand{\ang}{\rm \AA}
\newcommand{\msun}{M$_\odot$}
\newcommand{\lsun}{L$_\odot$}
\newcommand{\days}{$d$}
\newcommand{\degree}{$^\circ$}
\newcommand{\ud}{{\rm d}}
\newcommand{\as}[2]{$#1''\,\hspace{-1.7mm}.\hspace{.0mm}#2$}
\newcommand{\am}[2]{$#1'\,\hspace{-1.7mm}.\hspace{.0mm}#2$}
\newcommand{\ad}[2]{$#1^{\circ}\,\hspace{-1.7mm}.\hspace{.0mm}#2$}
\newcommand{\lsim}{~\rlap{$<$}{\lower 1.0ex\hbox{$\sim$}}}
\newcommand{\gsim}{~\rlap{$>$}{\lower 1.0ex\hbox{$\sim$}}}
\newcommand{\HA}{H$\alpha$}
\newcommand{\HII}{\mbox{H\,{\sc ii}}}
\newcommand{\kms}{\mbox{km s$^{-1}$}}
\newcommand{\HI}{\mbox{H\,{\sc i}}}
\newcommand{\KI}{\mbox{K\,{\sc i}}}
\newcommand{\nan}{Nan\c{c}ay}
\newcommand{\galex}{{\it GALEX}}
\newcommand{\jks}{Jy~km~s$^{-1}$}
\slugcomment{\it\footnotesize Submitted to ApJ March 16, 2015; revised May
  26; accepted
  June 9}

\title{New Measurements of the Radio Photosphere of Mira based on Data
  from the JVLA and ALMA}

\author{L. D. Matthews\altaffilmark{1},  M. J. Reid\altaffilmark{2}, \&
K. M. Menten\altaffilmark{3}}

\altaffiltext{1}{MIT Haystack Observatory, Off Route 40, Westford, MA
  01886 USA; lmatthew@haystack.mit.edu}
\altaffiltext{2}{Harvard-Smithsonian Center for Astrophysics, 60
  Garden Street, MS-42, Cambridge, MA 02138 USA}
\altaffiltext{3}{Max-Planck-Institut f\"ur Radioastronomie, Auf dem
  H\"ugel 69, D-53121 Bonn, Germany}

\begin{abstract}
We present new measurements of the millimeter wavelength continuum emission from the long period
variable Mira ($o$~Ceti) at frequencies of 46~GHz, 96~GHz, and 229~GHz
($\lambda\approx$7~mm, 3~mm, and 1~mm) based on
observations obtained with the Jansky Very Large
Array (JVLA) and the Atacama
Large Millimeter/submillimeter Array (ALMA). 
The measured millimeter flux densities are consistent
with a radio photosphere model derived from previous observations,
where flux density $S_{\nu}\propto\nu^{1.86}$. The
stellar disk is resolved, and the
measurements indicate a decrease in the size
of the radio photosphere at higher frequencies, as expected if the
opacity decreases at shorter wavelengths. The shape of the radio
photosphere is found to be slightly elongated, with a flattening of $\sim$10-20\%.
 The data also reveal evidence for brightness non-uniformities on the surface of Mira at
radio wavelengths. Mira's hot companion, Mira~B was detected at
all three observed wavelengths, and we measure a radius for its radio-emitting
surface of $\approx2.0\times10^{13}$~cm. 
The data presented here
highlight the power of the JVLA and ALMA for the study of the
atmospheres of evolved stars.

\end{abstract}


\section{Introduction\protect\label{intro}}
Mira ($o$~Ceti) is the archetype of its class of long-period variable
stars and is one of the most extensively studied stars in the
Galaxy.  Mira has an oxygen-rich
chemistry and is undergoing mass loss at a rate of
$\sim2\times10^{-7}~M_{\odot}$ yr$^{-1}$ (Ryde \& Sch\"oier
2001).\footnote{Throughout 
the paper we adopt a distance
  of 110$\pm$9~pc (Haniff et al. 1995), based on the period-luminosity
  relation of Feast et al. 1989.} The pulsation period of Mira is 332
days, during which the visible light of the star varies by up to $\sim$8 magnitudes.
Mira is also a member of the nearest symbiotic binary system, Mira~AB.
The companion, Mira~B, has a projected separation
of $\sim$\as{0}{5} (55~AU; Karovska et al. 1997; Ramstedt et al. 2014) and is likely
to be a white dwarf (Sokoloski \& Bildsten 2010).

Both components of the Mira~AB binary are detectable at radio wavelengths
(Matthews \& Karovska 2006). Matthews \& Karovska showed that the radio emission from
Mira~B is consistent with free-free emission originating from an
ionized circumstellar region $\sim10^{13}-10^{14}$~cm in
radius. Based on
measurements of the flux density $S_{\rm B,\nu}$ between 8.5 and 43.3~GHz,
they derived a fit to the radio spectrum of the form $S_{\rm
  B,\nu}=(0.010\pm0.008)\nu^{1.18\pm0.28}_{\rm GHz}$~mJy. 

The radio emission from Mira~A has an origin quite different from its companion. Reid
\& Menten (1997; hereafter RM97) showed that the centimeter wavelength emission
detected from
Mira~A and other nearby, 
long-period variables can be explained by the existence of a ``radio photosphere''
with a radius approximately twice that of the classical photospheric
radius $R_{\star}$ (defined by the line-free regions of the optical-infrared
spectrum). For Mira-type variables, $R_{\star}$ is typically
$\sim$1--2~AU, while between $\sim$1--$2R_{\star}$ is a molecular
layer (Perrin et al. 2004)
that may be nearly opaque in the visible and infrared. The radio
photosphere thus lies adjacent to this molecular layer.  

The emission arising from the
radio photosphere has a spectral
index $\alpha\approx$2 (consistent with optically thick blackbody
emission) all the way from centimeter to submillimeter
wavelengths. However, the 
flux densities over this spectral range are roughly a factor of two
higher than expected from an extrapolation of the optical photospheric
emission. Based on the model developed by RM97,
the dominant source of opacity  in the radio photosphere is
the interaction between free electrons
(mostly from ionization of K and Na) and neutral H and H$_{2}$. At
millimeter
and centimeter wavelengths, unity optical depth is achieved at a radius of
$\sim$3 to 4~AU, where the density and temperature
are $\sim10^{12}$~cm$^{-3}$ and $\approx1600$~K, respectively.

The detection and characterization of radio photospheres in asymptotic
giant branch (AGB) stars
provide a new tool for the investigation of the critical region
between the classic photosphere and the dust formation zone where the
stellar wind is
launched. In oxygen-rich AGB stars, the radio photosphere also lies just
interior to the atmospheric
region where SiO masers arise, and contemporaneous observations of these
maser lines provide additional diagnostics of the density, temperature,
dynamics, and shock propagation
within this complex and important atmospheric region (RM97; Gray et al. 2009).
As noted by RM97, red
giants with well-characterized radio photospheres may also be useful as
calibration sources in the millimeter and submillimeter regimes.

The characteristic sizes and temperatures of radio photospheres
imply that 
resolved imaging observations are within the
reach of current radio interferometers for nearby stars. Using the
Very Large Array (VLA) of the National Radio Astronomy Observatory\footnote{The National Radio Astronomy
  Observatory (NRAO) is a facility of the National Science Foundation
  operated  by Associated
  Universities, Inc. under cooperative agreement with the National
  Science Foundation.} in its highest resolution configuration, Reid \&
Menten (2007) obtained resolved images of three oxygen-rich
long-period variables (including Mira) at a wavelength of 7~mm,
while Menten et al. (2012) 
resolved the carbon star IRC+10216. One intriguing
finding from these studies is evidence that the radio photospheres
in some cases appear distinctly non-circular. However, since
observations were available at only a single epoch and at a single
point during the stellar pulsation cycle, it was difficult to
distinguish between possible origins for these observed shapes, and it
is unknown whether the shapes are static or time-variable.

The exquisite continuum sensitivity of the recently upgraded Jansky
VLA (JVLA; Perley et al. 2011) and the Atacama Large
Millimeter/submillimeter Array (ALMA; Wootten \& Thompson 2009) 
now provide a tremendous leap in
our ability to obtain high-quality images of radio photospheres over a
broad range of wavelengths. At 7~mm, the continuum sensitivity of the
JVLA exceeds that of the pre-upgrade VLA by an order of magnitude. At the same time ALMA
supplies sub-mJy sensitivity with only modest integration
times, and its longest baselines are capable of providing angular
resolution comparable to, or higher than the JVLA, depending on observing
frequency. Here we present an analysis of new 
observations of Mira obtained with the JVLA and ALMA 
at wavelengths of 7~mm (46~GHz) 3~mm (96~GHz), and 1~mm (229~GHz).
The quality of the data underscore the potential of these two
instruments for the study of stellar atmospheres. All of the data were obtained during
the course of a nine-month period (i.e., during a timescale less than
a single stellar pulsation
cycle) in 2014. 

\section{Observations\protect\label{observations}}
\subsection{JVLA Data}
Observations of Mira at 46~GHz (7~mm) were carried out with
the JVLA  in its most extended (A) configuration (0.68~km
to 36.4~km
baselines) on 2014 February 23 during a 
3-hour observing session. Data were recorded with 2-second integration times.
Offset corrections as refinements to the reference antenna pointing
model were evaluated hourly
using observations of a strong point source at 8~GHz. 
3C48 was observed as an absolute flux calibrator, and the bright continuum
source J0359+5057 (NRAO150) 
was observed as a bandpass calibrator. 

To obtain maximum continuum sensitivity, a 3-bit observing
mode was employed, and the WIDAR correlator was configured  
with 4 baseband pairs tuned to contiguously cover
the frequency range from 41.8-49.4 GHz with dual circular polarizations. 
Each baseband pair contained 16
subbands, each with a bandwidth of 128~MHz and 128 spectral channels.
This frequency range included the SiO $v$=1 and $v$=2,
$J$=1-0 maser transitions (see below).  The total on-source
integration time for Mira was approximately 1 hour 51 minutes, split
into two  observing blocks that were bracketed by observations of the gain
(amplitude and phase) calibrator J0215-0222.

\subsection{ALMA Data}
ALMA observations of Mira
were performed in Band~3 (94~GHz) on 2014 October 17 and 25  and in 
Band~6 (229~GHz) on 
2014 October 29 and 2014 November 1, respectively using an array 
configuration with baselines up to 15.2~km. The data were obtained by
ALMA staff for science verification purposes (ALMA Partnership 2015) 
and were released to the public
in 2015 February. A total of 38 (39) antennas were present in the array for
the Band~3 (6) observations.
The total on-source integration times for Mira were approximately 1.6 hours in Band~3 and 1.1
hours in Band~6. Data from the different observing dates were combined
for each band in the present analysis.

\section{Data Processing\protect\label{reduction}}
\subsection{JVLA Data\protect\label{jvlareduction}}
Data processing for the JVLA data was performed using the Astronomical Image Processing
System (AIPS; Greisen 2003), taking into account special
considerations necessary for handling JVLA data
as outlined in Appendix~E of the
AIPS Cookbook.\footnote{http://www.aips.nrao.edu/cook.html}. 

The data were loaded into AIPS directly from archival
science data model (ASDM) format
files using the  Obit software package (Cotton
2008). The default calibration (`CL') table was then
recreated  to  update the single-dish antenna gain curve and
atmospheric opacity corrections.

After updating the antenna positions and
flagging obviously corrupted data, a requantizer gain correction was
applied, as required for 3-bit data. Next, a fringe fit
  was performed using one minute of 
data from the observation of J0359+5057 to correct the
  instrumental delay settings. Separate delay solutions were
  determined for each of the four independent basebands.
Bandpass calibration was performed using NRAO150, assuming a 
spectral index $\alpha$=0.4 (Trippe et al. 2010).
The absolute flux density scale was
calculated using the latest time-dependent flux density values for
3C48 from Perley \& Butler (2013). The resulting flux 
density $S_{\nu}$ as a function of frequency
for 3C48
is therefore given by ${\rm log}(S_{\nu}) = 1.3324 - 0.7690({\rm
  log}(\nu)) - 0.1950({\rm log}(\nu))^{2} +0.059({\rm log}(\nu))^3$, where
  $\nu$ is the frequency expressed in GHz.
The frequency-independent portion of the
complex gains (amplitude and phase)
was calibrated using J0215-0222 to 
remove possible slow (hour timescale) instrumental gain
drifts. Additionally, a correction of $-3.324''$ in DEC
was applied to the observed (J2000.0) 
position of Mira to account for the large proper 
motion of the star. This was required
after the discovery that corrections for parallax and proper motion were
inadvertently excluded from the source position entered in the
observation scheduling
file.

Mira is a well-known source of SiO maser emission (e.g., Cotton et
al. 2006; McIntosh \&
Bonde 2010). We therefore employed the method developed by Reid \& Menten
(1990) to improve the phase calibration using the stellar maser signal. 
After completion of the initial calibration steps described above, the spectral channel
containing the brightest SiO $v$=1, $J$=1-0 maser emission was split 
from the main data set, and several iterations of phase-only
self-calibration were performed on the maser data 
until convergence was reached. From
these solutions, phase corrections appropriate for each of the 64
subbands comprising the full 8~GHz bandwidth of the continuum
measurements were derived using the AIPS task {\small\sc{SNP2D}}. 
(Effectively, the phase correction $\phi_{\nu'}$ for a given subband with
central frequency $\nu'$ then becomes 
$\phi_{\nu'}=(\nu'/\nu_{0})\phi_{0}$, where $\nu_{0}$
is the frequency of the reference channel and $\phi_{0}$ is the
computed phase correction in that channel). If the maser emission is
sufficiently strong, in principle this approach 
allows achievement of nearly perfect ``seeing'' in the resulting
continuum measurements (Reid \& Menten 1990; RM97). However, at the
phase Mira was observed, the SiO
masers were in a relatively low state (integrated flux density
$\sim$2.7~Jy), and the modest signal-to-noise ratio on some of the longest
baselines was a limiting factor in the calibration solutions (see Appendix).

After applying the derived phase corrections, a second round of
self-calibration was performed on the reference channel, this time
solving for both the amplitudes and the phases.  To mitigate possible
drift in the amplitudes, the gains were normalized during this step, and the
resulting change in the amplitude scale was measured to be less than 0.5\%.  Once these
gain corrections had been applied to the full data set, 
spectral channels 1-3 and 125-128 of each subband (where
noise is significantly higher) were flagged, along with spectral
channels containing line emission.  Next, the data were
averaged to a time resolution of 10 seconds to reduce the data volume
and Hanning smoothed to reduce 
ringing caused by the narrow maser lines.
At this stage, optimized weights for the visibility data were
computed using the AIPS task {\small\sc{REWAY}}, and finally, the data were
further averaged in frequency, resulting in 8 spectral channels in each of the
64 subbands (IFs).

\subsection{ALMA Data}
We retrieved the pipeline calibrated $u$-$v$ data in the form of 
CASA measurement sets for each of the two ALMA bands from the ALMA
Science Verification Data web pages.\footnote{https://almascience.nrao.edu/alma-data/science-verification.}
Details of the processing steps used to produce these data sets are
available at that site. 

The flux density scale for the Band~3 data
taken on 17 October was
calibrated using the source J0334-4008, assuming a flux density at
86.2251~GHz of 1.6589~Jy and a spectral index $\alpha=-0.7146$. For
the Band~3 data from 25 October, J0238+1636 and J2258-2758 were used as flux
density calibrators, assuming respectively $S_{\nu}=$1.8550~Jy,
$\alpha=-0.269071$ and $S_{\nu}$=1.1315~Jy,
$\alpha=-0.7282$. For Band~6,
the flux density scale for the 29 October data was set using
J0334-4008, assuming $S_{\nu}$=0.8282~Jy at 229.5497~GHz and
  $\alpha=-0.7029$. For the 01 November data, the flux calibrator was
J0238+1636, adopting $S_{\nu}$=1.3893~Jy and $\alpha=-0.2005$
at 229.5485~GHz.

At this stage, the continuum portion of the
Band~3 data included a total of nine spectral windows (three in
each of three independent basebands) and dual linear polarizations.
Each spectral window contained
eight spectral channels (250~MHz each) across a $\sim$1.8~GHz band
(i.e., with slight spectral overlap between channels). 
Adjacent channels thus have slight spectral overlap. 
The frequency ranges sampled were
87.3--89.0~GHz, 97.4--99.1~GHz, and 99.3--101.0~GHz.
The continuum portion of the Band~6 data contained two spectral windows (both within a single
baseband), again with dual linear polarizations and eight 250~MHz spectral channels  
across a $\sim$1.8~GHz total band. The frequency range covered by the
Band~6 observations was $\sim$228.7--230.4~GHz.

Using the continuum emission from the Mira~AB binary itself, we performed
self-calibration on the pipeline data for both ALMA
bands. Initially, several iterations of phase-only self-calibration
were performed using a solution interval equal to the record length of
the data ($\sim$6 seconds in Band~3 and $\sim$3 seconds in
Band~6) until convergence was reached. These solutions were applied, and additional iterations were
performed, solving for both amplitude and phase with a solution
interval of 60 seconds. To avoid a possible
drift in the amplitudes, the gains were normalized during this latter step, and the
resulting change in the amplitude scales of both bands were measured to be $<$0.7\%.

\subsection{Imaging the Data}
\begin{figure}
\vspace{-1.5cm}
\centering
\scalebox{0.5}{\rotatebox{0}{\includegraphics{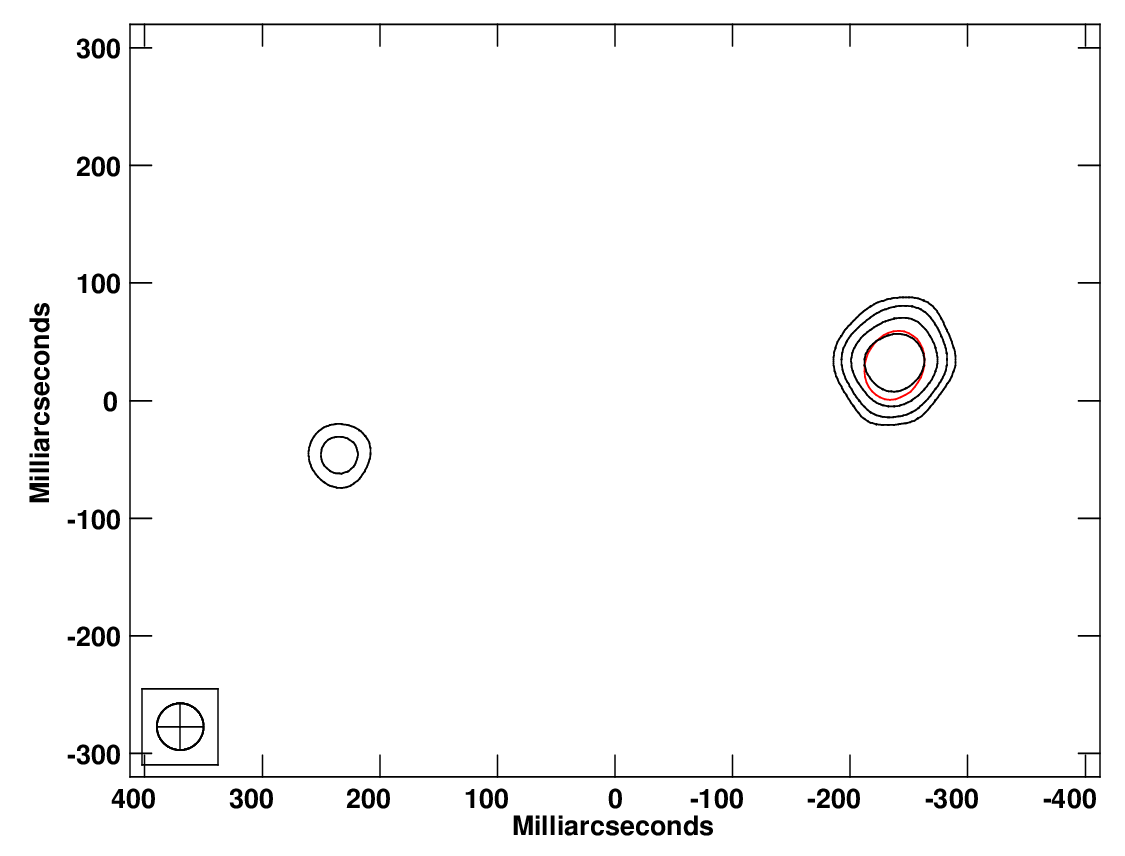}}}
\caption{Image of Mira~AB obtained with the JVLA at a frequency of
  46~GHz on 2014 February 23. 
Contour levels are ($-1$ [absent],1,2,4,8)$\times$290~$\mu$Jy
  beam$^{-1}$. The lowest contour is 10$\sigma$. The 
  restoring beam had a FWHM 39.7~mas. The red ellipse indicates the
  dimensions of the best uniform elliptical disk fit to the visibility data for Mira~A.  }
\label{fig:Qimage}
\end{figure}

\begin{figure}
\vspace{-1.5cm}
\centering
\scalebox{0.5}{\rotatebox{0}{\includegraphics{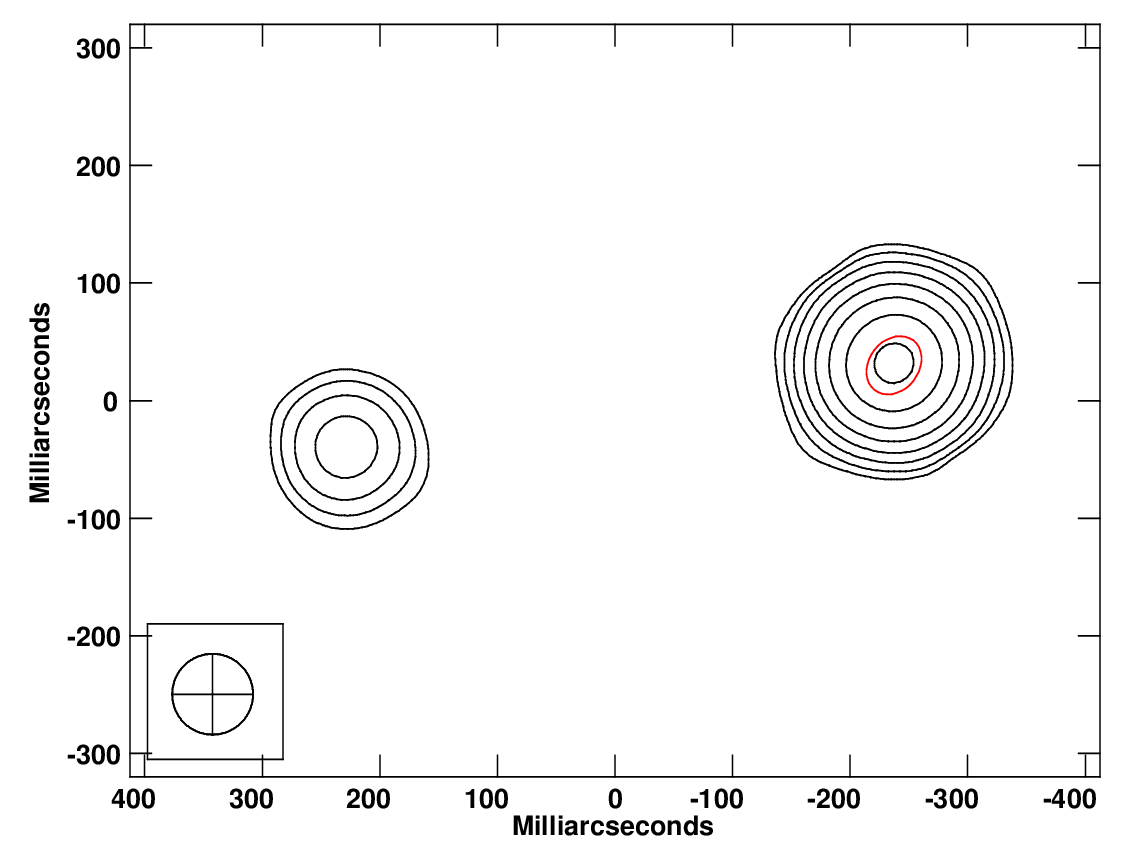}}}
\caption{Image of Mira~AB obtained from ALMA data at a frequency of
  94~GHz. 
Contour levels are ($-1$ [absent],1,2,4...128)$\times$197~$\mu$Jy
  beam$^{-1}$. The lowest contour is 10$\sigma$. The 
  restoring beam had a FWHM of 68.7~mas. The red ellipse indicates the
  dimensions of the best uniform elliptical disk fit to the visibility data for Mira~A.  }
\label{fig:B3image}
\end{figure}

\begin{figure}
\vspace{-1.5cm}
\centering
\scalebox{0.5}{\rotatebox{0}{\includegraphics{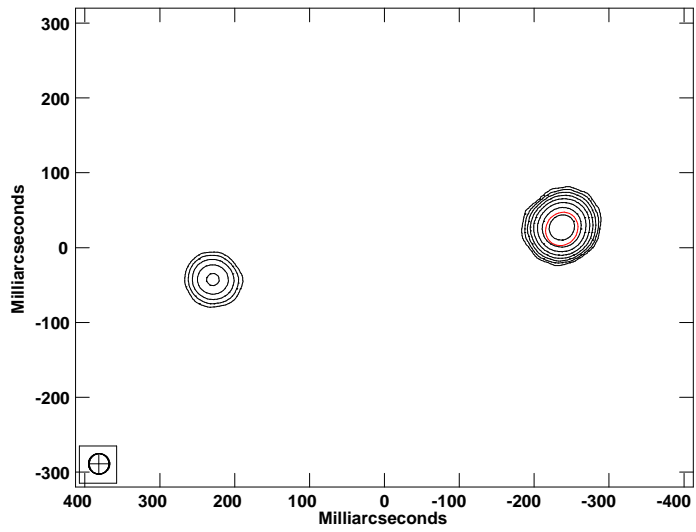}}}
\caption{Image of Mira~AB obtained from ALMA data at a frequency of
  229~GHz. 
Contour levels are ($-1$ [absent],1,2,4...128)$\times$348~$\mu$Jy
  beam$^{-1}$. The lowest contour is 10$\sigma$. The 
  restoring beam had a FWHM of 27.3~mas. The red ellipse indicates the
  dimensions of the best uniform elliptical disk fit to the visibility data for Mira~A.  }
\label{fig:B6image}
\end{figure}

Imaging and deconvolution of the JVLA and the ALMA 
data were 
performed using {\small\sc{IMAGR}} in AIPS. For the images presented
here, we used a Briggs robust value of ${\cal R}$=0, a cell size of
5~mas, and a circular restoring beam
with a FWHM equal to the mean of the dimensions of the dirty
beam at the appropriate frequency (see Table~1). Corrections were applied during imaging for the
frequency dependence of the primary beam and the expected spectral
index of Mira~A within the JVLA band ($\alpha$=1.86; RM97).  Images of the Mira~AB binary
at each of the three observed frequencies are presented in
Figures~\ref{fig:Qimage}, \ref{fig:B3image}, and \ref{fig:B6image}.


%
\begin{deluxetable}{lccccc}
\tabletypesize{\tiny}
\tablewidth{0pc}
\tablenum{1}
\tablecaption{Deconvolved Image Characteristics}
\tablehead{
\colhead{Band} & \colhead{$\theta_{\rm B}$(maj)} & \colhead{$\theta_{\rm B}$(min)} & 
\colhead{PA} & \colhead{$\theta_{\rm R}$} & \colhead{rms noise}\\
\colhead{}     & \colhead{(mas)}                 & \colhead{(mas)} &
\colhead{(deg)} & \colhead{(mas)} & \colhead{($\mu$Jy beam$^{-1}$)}\\
\colhead{(1)}     & \colhead{(2)}                 & \colhead{(3)} &
\colhead{(4)} & \colhead{(5)} & \colhead{(6)}
}

\startdata
Q (46~GHz) & 45 & 34 & $-23$ & 39.7 & 29.0\\

3 (96 GHz) & 79 & 58 & +42 & 68.7 & 19.7 \\

6 (229 GHz) & 32 & 23 & +19 & 27.3 & 34.8\\

\enddata

\tablecomments{$\theta_{\rm B}$(maj) and $\theta_{\rm B}$(min) are the
  major and minor axis diameters of the synthesized dirty
  beam measured at
  FWHM using ``robust 0'' weighting of the visibilities. 
The position angle (PA) of the beam is measured east from north. 
Deconvolved images used for the present analysis
  used circular restoring beams with FWHM as specified in column~5,
  and where $\theta_{R}$ is equal to the mean of $\theta_{\rm B}$(maj) and $\theta_{\rm B}$(min).}

\end{deluxetable}


%
\begin{deluxetable}{lcllllll}
\tabletypesize{\tiny}
\tablewidth{0pc}
\tablenum{2}
\tablecaption{Measured Photospheric Parameters for Mira A}
\tablehead{
\colhead{Band} & \colhead{Stellar Phase} & \colhead{$\theta_{\rm
    maj}$} & \colhead{$\theta_{\rm min}$} &
\colhead{PA} & \colhead{$\epsilon$} & \colhead{$S_{\nu}$} & \colhead{$T_{b}$}\\
\colhead{}  & \colhead{} & \colhead{(mas)} & \colhead{(mas)} & \colhead{(degrees)}  
& \colhead{} & \colhead{(mJy)}    & \colhead{(K)}\\
  \colhead{(1)} & \colhead{(2)} & \colhead{(3)} & \colhead{(4)} & \colhead{(5)}
& \colhead{(6)} & \colhead{(7)} & \colhead{(8)} }

\startdata

\tableline
\\
\multicolumn{8}{c}{Values Based on Uniform Elliptical Disk Fits to Visibility Data}\\
\\
\tableline

Q (46~GHz) & 0.76 & 60.0$\pm$2.0 (0.7) & 49.3$\pm$1.6 (0.5) & $156\pm$4 (1) & 0.82$\pm$0.04 &
7.4$\pm$1.5 (0.04) &  2110$\pm$440 \\

3 (96~GHz) & 0.47 & 53.4$\pm$11.2 (0.1) & 42.3$\pm$13.0 (0.1) &
142$\pm$36 (0.3) &
0.79$\pm$0.29 & 34.4$\pm$6.9 (0.01) & 2900$\pm$1200\\

6 (229~GHz) & 0.51 & 46.2$\pm$2.3 (0.03) & 41.5$\pm$2.3 (0.02) &
144$\pm$11 (0.1) &
0.90$\pm$0.07 & 150$\pm$30 (0.04) & 2620$\pm$560\\

\tableline
\\
\multicolumn{8}{c}{Values Based on Elliptical Gaussian Fits to Image Data}\\
\\
\tableline
Q (46~GHz) & 0.76 & 37.5$\pm$2.1 (0.7) & 31.7$\pm$2.0 (0.7) &
147$\pm$6 (4) & 0.84$\pm$0.07 & 7.6$\pm$1.5 (0.07) &
...\\

3 (96 GHz) & 0.47 & 32.8$\pm$18.3 (0.2) & 26$\pm$26(0.2) &
139$\pm$3 (1) & 0.8$^{+0.2}_{-0.8}$ & 34.4$\pm$7.0 (0.04) &
...\\

6 (229 GHz) & 0.51 & 30.4$\pm$0.4 (0.03) & 27.0$\pm$0.4 (0.03) &
143.4$\pm$0.5 (0.4) & 0.89$\pm$0.02 &
153$\pm$31 (0.09) & ...\\

\enddata
\tablecomments{Stellar phase was derived using the ephemeris available from
 the database of the
  American Association of Variable Star Observers (AAVSO) at
  http://www.aavso.org. The error bars on the measured quantities include formal,
  systematic, and calibration uncertainties, as
described in the Appendix.
For measured quantities, the value in
parentheses indicates the contribution to the total error budget from
the formal fit uncertainties.
 For a uniform disk,
$\theta_{\rm maj}$ and $\theta_{\min}$ are the major and minor axis sizes of
the disk, respectively; for a Gaussian fit they represent the
the FWHM dimensions of the elliptical Gaussian, after deconvolution with the size
of the dirty
beam. For a resolved source, $\theta_{\rm maj}$ measured from a  Gaussian fit is
expected to be 
0.625 times the value of $\theta_{\rm maj}$ measured from a uniform
  disk fit.  The ellipticity, $\epsilon$, is calculated as the ratio of
  $\theta_{\rm min}$ to $\theta_{\rm maj}$.  The brightness temperature was
  computed as $T_{b}=2S_{\nu}c^2/(k\nu^{2}\pi\theta_{\rm
    maj}\theta_{\rm min})$ where $c$ is the speed of light and $k$ is
  the Boltzmann constant.   }

\end{deluxetable}

%
\begin{deluxetable}{lllll}
\tabletypesize{\tiny}
\tablewidth{0pc}
\tablenum{3}
\tablecaption{Measured Parameters for Mira B}
\tablehead{
\colhead{Band}  & \colhead{$\theta_{\rm
    maj}$} & \colhead{$\theta_{\rm min}$} &
\colhead{PA}  & \colhead{$S_{\nu}$} \\
 \colhead{} & \colhead{(mas)} & \colhead{(mas)} & \colhead{(degrees)}  
& \colhead{(mJy)}   \\
  \colhead{(1)} & \colhead{(2)} & \colhead{(3)} & \colhead{(4)} & \colhead{(5)}
  }

\startdata

Q (46 GHz) & 18$\pm$5 (4) & 11$\pm$6 (5) & 172$\pm$30 (29) & 0.97$\pm$0.20 (0.05)\\

3 (96 GHz)$^{a}$ & 20$\pm$20 (3) & 19$\pm$19 (4) & 46$\pm$32 (32) &
2.5$\pm$0.5 (0.04) \\

6 (229 GHz) & 25.4$\pm$0.6 (0.4) & 23.4$\pm$0.6 (0.4) & 69$\pm$7 (7) &
11.5$\pm$2.3 (0.09)\\

\enddata

\tablenotetext{a}{After accounting for systematic errors, 
Mira~B is formally indistinguishable from a point  source in the 
Band~3 data. }
\tablecomments{Parameters are derived from elliptical Gaussian fits to 
the image data described in
  Table~1. The error bars on the measured quantities include formal,
  systematic, and calibration uncertainties, as
described in the Appendix. For measured quantities, the value in
parentheses indicates the contribution to the total error budget from
formal fit uncertainties. }

\end{deluxetable}

\section{Results}
\subsection{The Millimeter Wavelength Flux Densities and Sizes of Mira A and B}
\subsubsection{Mira A\protect\label{miraAshapesize}}
We have used two approaches to measure the flux density and size
of Mira~A based on the current data. The first was to fit
a uniform elliptical disk model to the visibility data at each observed
frequency using the AIPS task {\small \sc{OMFIT}}.   The results are
summarized in the top section of Table~2. The second method was to fit elliptical
Gaussians in the image plane using the ``robust 0''
images summarized in Table~1. 
General expressions for an elliptical Gaussian and for the visibility
amplitude of a uniform disk may be found in Condon (1997) and Pearson
(1999), respectively. For a resolved source, 
the FWHM size measured from an elliptical  Gaussian fit (after
deconvolution with the size of the dirty beam) is
expected to be a factor of
0.625 times the major axis of the elliptical disk
fit (Pearson 1999), consistent with our measurements. The derivation
of the error bars given in Table~2 is described in the Appendix.

The sizes derived from our uniform elliptical disk fits are overlaid
on the contour 
images presented in Figures~\ref{fig:Qimage}, \ref{fig:B3image}, and \ref{fig:B6image}. 
We have resolved the stellar disk at
all three of the observed frequencies. 
 Furthermore,
our measurements point to a trend of decreasing size for the radio
photosphere with increasing frequency.  The arithmetic means of the 
angular diameters in Table~2 translate to radio
photospheric radii of 4.5$\times10^{13}$ cm (3.0~AU) at 46~GHz,
3.9$\times10^{13}$~cm (2.6~AU) at 96~GHz, and 3.6$\times10^{13}$~cm 
(2.4~AU) at 229~GHz.
The radio photosphere model of RM97 predicts that the size of the
radio photosphere should decrease at higher frequencies, and this is
the first time that this has been confirmed observationally for an AGB
star. A similar effect has been previously seen in the red
supergiant Betelgeuse (Lim et al. 1998).

The angular sizes that we measure for the radio photosphere are intermediate between
values previously measured for the ``molecular layer'' of Mira in the
infrared (Perrin
et al. 2004) and the sizes of the SiO $v$=1 and $v$=2, $J$=1-0
maser emitting ``rings'', whose radii have been found to range between $\sim$30--40~mas
(3.3-4.4~AU) over the course of the stellar pulsation cycle (Cotton et
al. 2006). These results suggest that the SiO masers arise just outside
the radio photosphere, consistent with previous findings (Reid \&
Menten 2007; see also Gray et al. 2009).
In addition to resolving the photosphere, our new measurements point
to a statistically significant elongation in the
shape of Mira's radio photosphere at a level of $\sim$10-20\% (i.e.,
$\epsilon\approx$0.8-0.9), with the major axis of the elongation lying 
along a position
angle (PA)
between 139 to 156 degrees. 

\begin{figure}
\centering
\scalebox{0.7}{\rotatebox{0}{\includegraphics{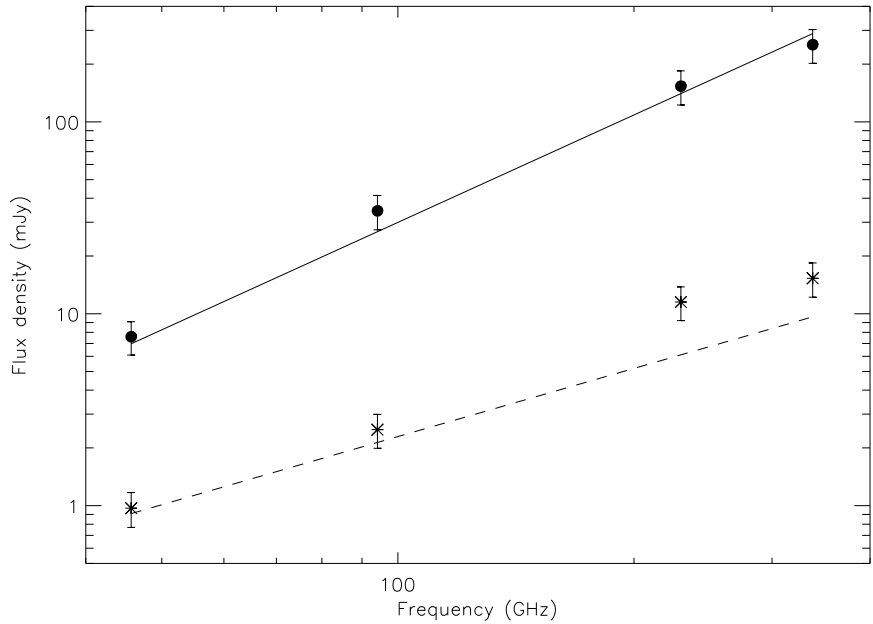}}}
\caption{Radio spectra for Mira A (solid points) and Mira~B
(asterisks). The three lower frequency points are from the present
  study and the point at 338~GHz is from Ramstedt et al. 2014. Flux
  densities were derived from fits to image data (Table~2). The solid line is the spectrum
predicted for Mira~A by the radio photosphere model of RM97 (Equation~2), and the dashed line
is the spectrum derived from earlier measurements of Mira~B 
(between 8-43~GHz) from the study of Matthews \&
Karovska 2006 (see \S\ref{intro}). }
\label{fig:spectrum}
\end{figure}

Figure~\ref{fig:spectrum} plots the short wavelength radio spectrum of Mira~A (and
Mira~B, 
discussed below) 
based on our current measurements. Additional data points at 
338~GHz from Ramstedt et al. (2014) are also
included. The latter measurements were derived from a combination of
ALMA data taken on
25 February 2014 and 2014 May 3 with a relatively compact antenna
configuration that only marginally resolved the two stars. 

The solid line overplotted on Figure~\ref{fig:spectrum} is not a fit
to the data, but rather represents the flux density as a function
of frequency predicted by the radio
photosphere model of RM97: 

\begin{equation}
S_{\nu}=0.50\left(\frac{\nu}{10 {\rm
      GHz}}\right)^{1.86}\left(\frac{D}{{\rm 100
      pc}}\right)^{-2}~{\rm mJy,}
\end{equation}

\noindent where $D$ is the distance to the star in parsecs. 
We see that this model provides
an excellent match to the data for our adopted distance of 110~pc.

Based on the parameters derived from our uniform disk fits, we present in
column~8 of Table~2 the brightness temperature, $T_{b}$, of Mira computed for
each of the three observed frequencies. Although the uncertainties in
the derived $T_{b}$ values are rather large, these values are
systematically higher than expected from the radio photosphere model of
RM97, which predicts brightness temperatures of $\approx$1600~K. Our values
are also higher than the measurement of
$T_{b}=1680\pm$250~K previously reported for Mira
by Reid \& Menten (2007). Temperatures higher than
those  predicted by the RM97 model
would imply an increase in opacity owing to more efficient collisional
ionization of K, Na, and other low-ionization elements. Consistent
with this, the photospheric radii
that we measure are systematically smaller than the RM97 model predicts
(ignoring uncertainties in the adopted stellar distance). 

The 
brightness temperature measurements in Table~2 represent lower limits to the stellar
effective temperature of Mira, and indeed, we find that they are 
comparable to, or smaller than most published 
effective temperature values [e.g., $T_{\rm
  eff}\sim2630-2800$~K (Haniff et al. 1995); $T_{\rm
    eff}\sim2918-3192$~K (Woodruff et al. 2004)].  However, the
  $T_{b}$ values systematically
  exceed the temperature of $\sim$2000~K derived for the molecular
  layer of Mira by Perrin et al. (2004). This is somewhat surprising in light
of the expectation that the radio photosphere should lie just exterior to
this molecular layer (RM97).

\subsubsection{Mira B}
Measurements of the millimeter wavelength flux density for Mira~B at each of the three observed
frequencies are presented in Table~3 and are plotted on the
spectrum in Figure~\ref{fig:spectrum}. 
We see from Figure~\ref{fig:spectrum} that
at 46~GHz and 96~GHz, the measured flux densities of Mira~B match closely with the
relation derived by Matthews \& Karovska (2006). However, both of the two
higher-frequency measurements lie significant above the relation. The origin of this
deviation is unclear. The model developed by
 Seaquist et al. (1984) and Taylor \& Seaquist (1984) for radio
emission from
symbiotic binaries
predicts a flattening or downturn in the flux
density at higher frequencies, opposite to the observed
trend. Furthermore, the
period of the Mira~AB binary  ($\sim$500~yr; Prieur et al. 2002)  is
too long for orbital modulation to have a measurable effect on the
spectral index over the course of the one-year period when these data
were obtained.

Based on the Gaussian fits presented in Table~3, it is clear that we have
resolved the radio surface of Mira~B at the highest observed
frequency. At 46~GHz, the source also appears to be marginally
resolved, although the inferred size is only a fraction of the
synthesized beam and the uncertainties are large. At 96~GHz, the
inferred size is consistent with being intermediate between the 
other two bands; however, after
consideration of systematic errors, Mira~B cannot  formally be
distinguished from a point source at this frequency.

Based on the ALMA 229~GHz data, we derive a radius for Mira~B of
$\approx2.0\times10^{13}$~cm. This is many orders of magnitude larger than the
expected photospheric radius of a white dwarf. This is therefore consistent with the radio
emission arising from thermal bremsstrahlung from an ionization-bounded
hypercompact \HII\
region surrounding Mira~B, embedded in Mira~A's neutral wind. However, for a canonical electron
temperature of 10,000~K, the measured
radius is approximately four times larger
than the radius that would be required for an optically thick blackbody to yield the
measured flux density. This, coupled with the apparent change in
spectral index at high frequencies imply that a more complex model is needed to
fully describe the observed short-wavelength radio emission. This is not surprising
given the complex environment surrounding the two stars. In
particular, Mira~A's circumstellar envelope is highly structured on
the scales that are relevant here (e.g., Mayer et al. 2011;
Ramstedt et al. 2014).

\subsection{Evidence for Non-uniformities on the Surface of Mira A}
\begin{figure}
\vspace{-1.5cm}
\centering
\scalebox{0.5}{\rotatebox{0}{\includegraphics{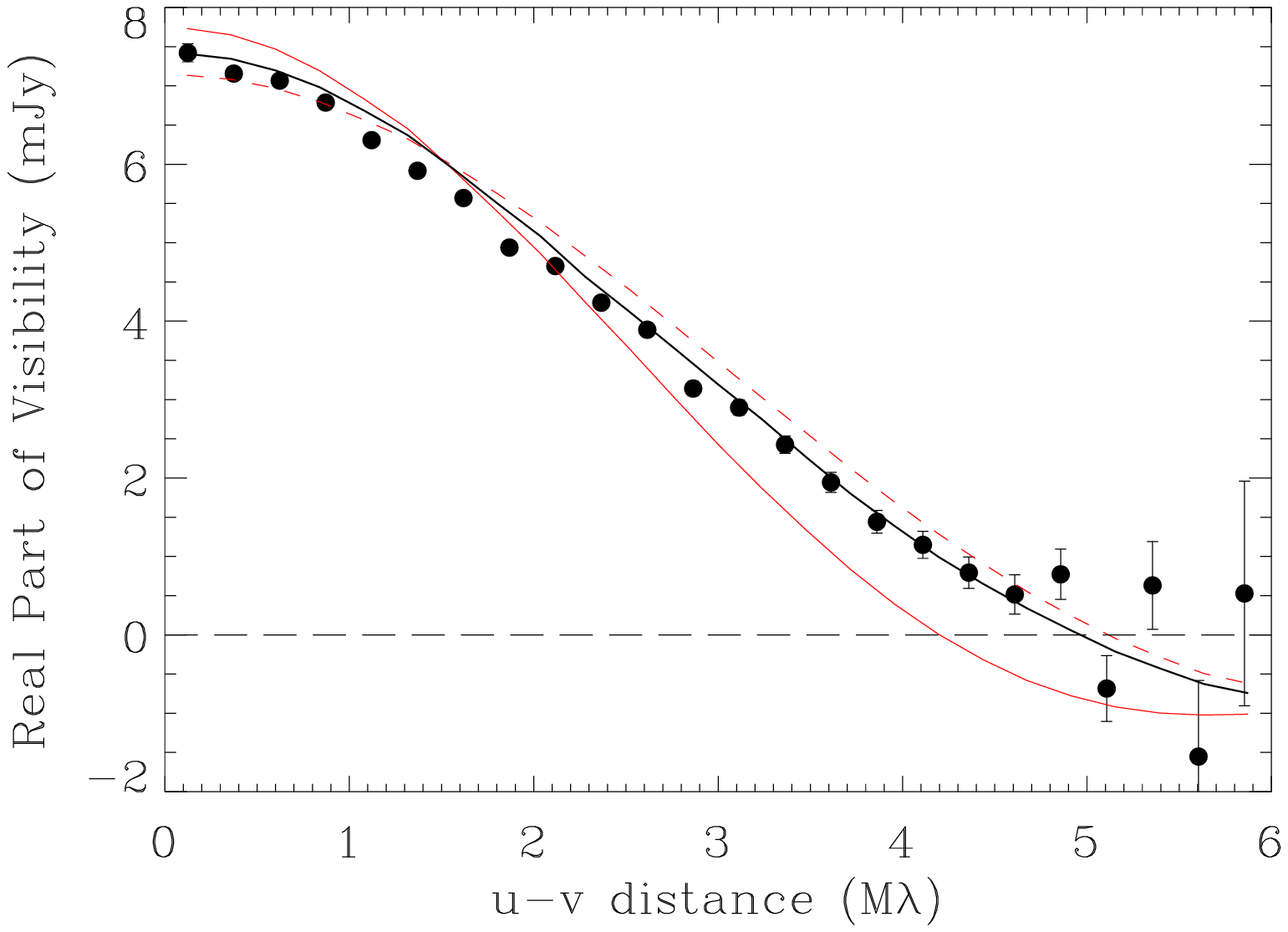}}}
\scalebox{0.5}{\rotatebox{0}{\includegraphics{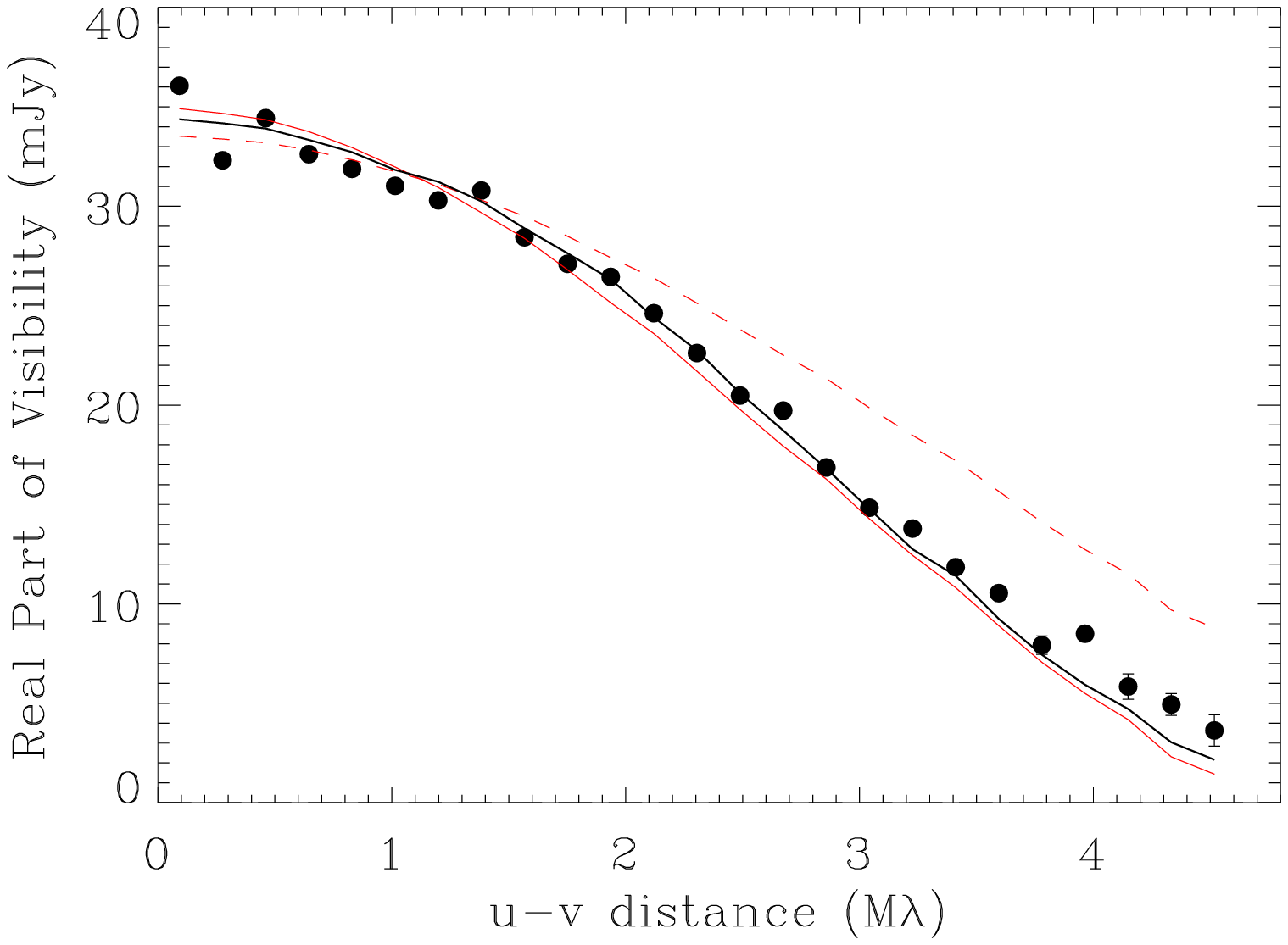}}}
\scalebox{0.5}{\rotatebox{0}{\includegraphics{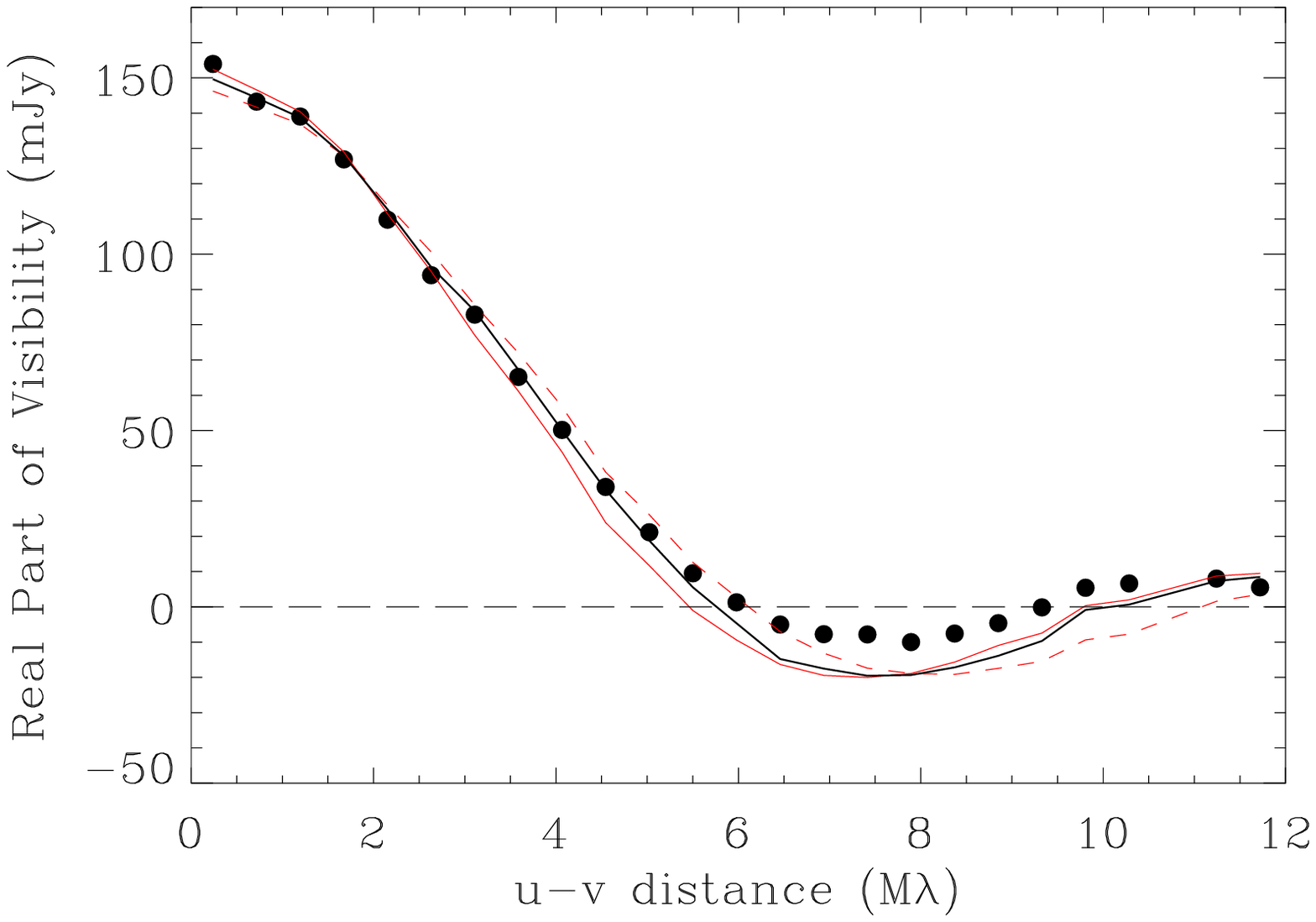}}}
\caption{Visibility versus baseline length for the three data sets
  presented in this paper: 46~GHz (top); 94~GHz (middle); 229~GHz
  (bottom). The solid black line shows the best-fitting uniform
  elliptical disk model from Table~2. 
The red lines represent circular uniform disk models with
 diameters equal to the major axes (solid lines) and minor axes (dashed
  lines) of the 
  elliptical disk fits.  }
\label{fig:uvplots}
\end{figure}

The high spatial resolution, high signal-to-noise ratio, and good
$u$-$v$ coverage of the three data sets presented here allow the most
detailed investigation to date of the brightness distribution on the
radio surface of Mira~A. In Figure~\ref{fig:uvplots} we present plots of the
azimuthally averaged
visibilities as a function of baseline length for the three observed frequencies.
Overplotted as thick black lines are the best-fitting uniform
elliptical disk models as presented in Table~2.
Models of circular uniform disks are overplotted in red. The solid red
line is circular model with a disk diameter equal to the major axis of the
best-fitting elliptical model, while the dashed line has a diameter
equal to the minor axis of the best-fitting elliptical model.

\begin{figure}
\vspace{-1.5cm}
\centering
\scalebox{0.7}{\rotatebox{0}{\includegraphics{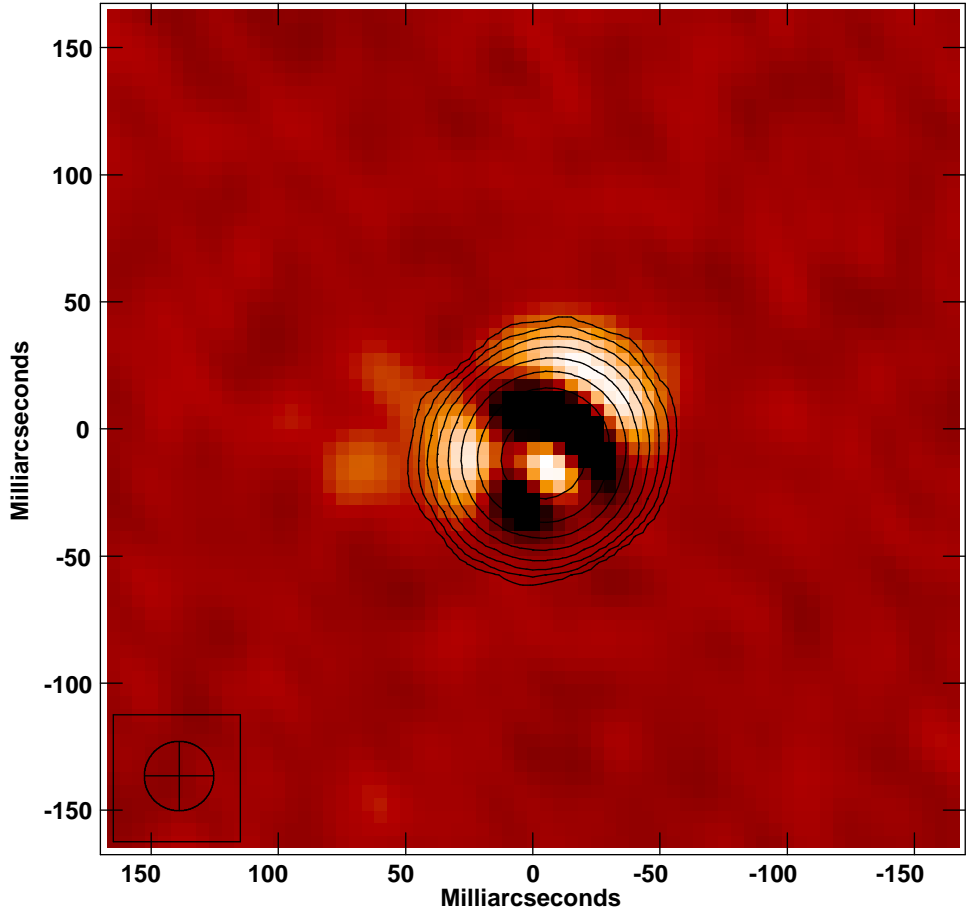}}}
\caption{Contour map of the 229~GHz emission from Mira~A (reproduced
  from Figure~\ref{fig:B6image}), overplotted on an image
  of the residuals after subtraction of the best-fitting uniform elliptical
  disk model. The residuals range from $-1.4$ to $+0.9$~mJy beam$^{-1}$, with
  the peak positive residual lying slightly southwest of the nominal
  disk center. In comparison, the peak 
  surface brightness of the observed stellar emission is 71.1~mJy beam$^{-1}$. The
restoring beam (FWHM 27.3 mas) is indicated in the lower left corner.}
\label{fig:B6residuals}
\end{figure}

We see that a simple uniform disk model does not provide a fully
adequate characterization of the data at any of the three observed
frequencies. At 46~GHz and 96~GHz, the uniform disk model slightly
over-predicts the measured visibility at $u$-$v$ spacings near
1-2~M$\lambda$ and  0.5-1~M$\lambda$, respectively, while
under-predicting the values on the longest baselines. Deviations from
the uniform elliptical disk model are also evident at 229~GHz, with the
model fitting poorly across the longest baselines and the deviations
showing a high level of significance. To see this effect in the image
plane, in Figure~\ref{fig:B6residuals} we plot the 229~GHz contours
for Mira~A over an image formed by subtracting the best-fitting
elliptical disk model
in the visibility domain and imaging the residuals. We also produced an 
analogous image
by differencing a deconvolved image of the real data with a deconvolved image of the best 
uniform disk model, and the result is virtually 
indistinguishable, suggesting that
the image residuals are not solely artifacts of the deconvolution
process or limited image fidelity. Contamination from a
spectral line within our band can also be excluded as a likely cause
of the apparent brightness non-uniformity, as oxygen-rich
red giants typically contain only a handful of weak molecular lines over the
observed bandwidth (Tenenbaum et al. 2010; Kami\'nski et al. 2013).

\begin{figure}
\vspace{-1.5cm}
\centering
\scalebox{0.7}{\rotatebox{0}{\includegraphics{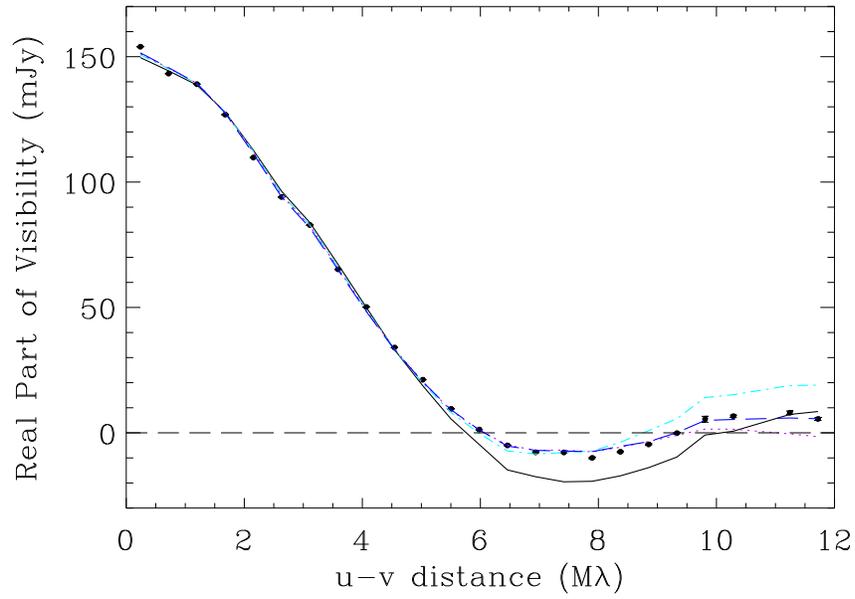}}}
\caption{Same as Figure~\ref{fig:uvplots}c, but with additional models
overplotted. The black solid line is the uniform elliptical disk model
from Table~1. The colored lines show models that include a uniform
elliptical disk plus an additional component: disk+point source (turquoise
dash-dot line
line); disk+ring (purple dotted line); disk+Gaussian (blue dashed line). }
\label{fig:B6visplotfits}
\end{figure}

We attempted to improve the fits to the visibility data using models
that include components in addition to the
uniform disk, including a ring (emulating limb brightening), a 
point source, or a Gaussian. For the 46~GHz and
96~GHz data, we were unable to meaningfully constrain models including
additional components, and none of these three simple models produced a
significant improvement in the fit (as assessed by visual inspection of
the residuals
and the improvement in the $\chi^{2}$ statistic). However, for the
229~GHz data, two-component models produced a clear
improvement in $\chi^{2}$, with the elliptical disk+Gaussian fit providing the best match to
the data (Figure~\ref{fig:B6visplotfits}). A fit with a 
disk plus two point sources was also
attempted, but was less satisfactory than the elliptical disk+Gaussian fit and
required an additional free parameter. 

For the elliptical disk+Gaussian model shown in
Figure~\ref{fig:B6visplotfits}, the position angle of the disk
component was fixed at 144.4$^{\circ}$ (i.e., to the value found
originally from fitting a disk model without additional components; see
Table~2). 
The total flux density in this
model was 151~mJy, with 39\% in a Gaussian component with FWHM 26~mas and a
minor-to-major axis ratio of 0.92, centered 3.5~mas southwest of the
nominal disk center, i.e., at the location of the bright region
visible in Figure~\ref{fig:B6residuals}. In this model, the major and
minor axes of the disk component are 13\% larger and 2\% smaller,
respectively, compared with the fit given in Table~2. 
While the elliptical disk+Gaussian model is not unique, 
its success nonetheless underscores that the observed
visibilities point to a nonuniform, asymmetric brightness
distribution. 

In its simplest form, the radio photosphere model of RM97 is not expected to exhibit a
temperature inversion with increasing height, although the dissipation of energy from shocks could
potentially lead to a region of increased temperature and level of
ionization just above the radio photosphere.  
Such a region is expected to become detectable at
 millimeter and submillimeter wavelengths and may manifest itself as detectable limb
brightening. While the present data do not rule out the presence of
a chromosphere-like region, the  fits presented in 
Figure~\ref{fig:B6visplotfits} suggest that models without azimuthally
symmetric limb
brightening are able to reproduce the data equally well, or better than
models that include it. 

\section{Discussion}
We have found excellent agreement between the newly observed
millimeter wavelength  flux
densities for Mira~A and the radio photosphere model of RM97
(Figure~\ref{fig:spectrum}), offering strong observational support for
this model and its extension to higher frequencies. However, we note that
our current measurement of the flux density of Mira~A at 46~GHz is
somewhat higher than previous measurements at comparable frequencies. 
Based on 43~GHz data taken on 2000 October~26, 
Reid \& Menten (2007) measured
$S_{\rm 43 GHz}$=4.8$\pm$0.2~mJy, while based on data from 2005, 
Matthews \& Karovska (2006)
reported values of 2.44$\pm$0.44~mJy (on May 17) and 2.61$\pm$0.37~mJy
(on June 28). Matthews \& Karovska did not have available simultaneous
measurements of the SiO maser to use for phase self-calibration (see
\S~\ref{jvlareduction}), and we estimate that decorrelation 
losses  resulting from the
larger residual phase scatter in their data
may have led to amplitude losses 
as high as $\sim30$\%. Nonetheless, even accounting for
this effect, as well as uncertainties in the absolute flux density
scale of up to 20\% (typical of measurements at this frequency), 
these data and the earlier
measurement of Reid \& Menten still appear to be consistent with a lower
43~GHz flux
density for Mira~A compared with our latest measurement. 

The
measurements of Reid \& Menten (2007) and  Matthews \& Karovska (2006) were all
obtained near a stellar phase $\phi\sim$0 (i.e., near maximum optical
light). Thus
while an evolution in the radio and millimeter flux density of Mira
over timescales of several years cannot be excluded, another
possibility is that the radio flux density varies over the course of the
stellar pulsation cycle. Indeed, previous 8.5~GHz 
measurements by RM97 at several phases during the
pulsation cycle of Mira implied
flux density variations of $\sim$15\% at that frequency and hinted at
a flux density minimum  near $\phi$=0.  RM97
showed that such a trend could be reproduced by a radio photosphere model
with low-level shocks (propagation speeds $\sim$10~\kms; see also Reid
\& Goldston 2002). 
Additional high-cadence monitoring
at radio and millimeter wavelengths is
clearly needed to settle the question of whether Mira's radio flux
density varies appreciably over the course of the stellar pulsation cycle,
over longer timescales (several years), or both. Further, simultaneous
monitoring at several frequencies, including longer wavelengths where
the stellar disk is unresolved, would help to distinguish
apparent brightness variations caused by 
uncertainties in the absolute flux density calibration of the JVLA and ALMA
at millimeter wavelengths. Calibration at these wavelengths,
particularly for long baseline measurements, is
notoriously difficult
since most strong, compact radio sources are highly variable at these
wavelengths, while planetary sources, despite having well determined
millimeter flux densities,
become resolved on long baselines (see RM97 for additional discussion).

Both the size and ellipticity that 
we measure for Mira's radio photosphere at 46~GHz are consistent 
with the values previously measured at 43~GHz by Reid \&
Menten (2007), although the PA measured by those authors
(+39$^{\circ}\pm50^{\circ}$) deviates from our more recent value. This hints at the
possibility that the shape of the photosphere may change with
time. However, given the large formal uncertainty
in the earlier PA measurement, this difference is only marginally
significant, and additional measurements will be needed to
confirm this possibility and to place  constraints on
the timescale of any such changes. New data sets of the quality that we present here would
also permit searches for temporal variations as small 
as $\sim$10\% in the photospheric
radius, even allowing for uncertainties caused by possible brightness non-uniformities in the
stellar emission. 
Previous measurements in the near- and mid-infrared have already
shown evidence of radius variations of this magnitude in Mira over the course of
the stellar pulsation cycle
($\sim$10\%; Weiner et al. 2003; Woodruff et al. 2004).

Measurements at other wavelengths have also supported the existence of
an elongation 
of Mira's photosphere comparable to the one we measure at millimeter 
wavelengths. For example, 
$\epsilon\approx$0.85
has been reported based on optical studies 
(Karovska et al. 1991; Wilson et al. 1992), although those
studies suggest that the position angle of the elongation varies with
time. Using {\it Hubble Space Telescope} 
ultraviolet imaging data, Karovska et al. (1997) also reported an
elongation along PA$\approx175^{\circ}$ (see also Karovska et
al. 2005), while
Tatebe et al. (2008) showed that  measurements 11.15$\mu$m  are consistent
with an ellipticity $\epsilon\sim0.9$ along PA$\approx$169$^{\circ}$.

There are several possible causes for non-circular shapes of radio
photospheres, including the presence of a close binary companion (see
Mauron et al. 2013), non-radial pulsations, rotation, or magnetic
fields. Our present data do not yet offer strong constraints on these
various possibilities, but this is expected to change as more
extensive databases of radio photosphere measurements become available
from the JVLA and ALMA in the coming years.

A simple calculation suggests that the tidal
force exerted by Mira~B would be insufficient to cause a distortion of
the observed magnitude. One may calculate an order of magnitude
estimate of the change in radius
or ``tidal bulge'', $\Delta R$,
raised by the presence of a perturbing potential along a line between
the center-of-mass of  the two bodies as:

\begin{equation}
\Delta R =
\frac{m_{2}}{m_{1}}\left(\frac{R}{s}\right)^{3}R
\end{equation}

\noindent where $m_{1}$ is the mass of the primary body, $m_{2}$ is
the mass of the perturber, $R$ is the radius of the star, and $s$ is the
separation between the two bodies (e.g., Danby 1992). Assuming
$m_{1}=2.5~M_{\odot}$ and $m_{2}=0.6~M_{\odot}$ (Prieur et al. 2002), 
$R=4\times10^{13}$~cm (\S\ref{miraAshapesize}), and
  $s$=100~AU (Prieur et al.), then $\Delta r\approx 2\times10^{8}$~cm. This implies an
  ellipticity $\epsilon\sim5\times10^{-6}$, several orders of
  magnitude smaller than observed. One possibility is that Mira has
  an additional unseen companion at a much smaller orbital
  radius. We
  also note that more sophisticated magnetohydrodynamic simulations have
  shown that a mechanism known as wind Roche-Lobe overflow may 
impose significant asymmetries on the shape of Mira's wind
  despite the large separation between the binary components (Mohamed \&
  Podsiadlowski 2012). In the simulations of Mohamed \&
  Podsiadlowski, distortion is seen in the dust shells as close as
  10~AU from the star, although this study does not specifically address the
  magnitude of any expected distortion on the radio photosphere, which
  lies interior to the wind launch region.

Both observations (e.g., Cotton et al. 2006) and theoretical modeling
(Thirumali \& Heyl 2013) point to magnetic fields as being important
in the atmospheric dynamics of Mira and in AGB stars in general (see
review by Vlemmings 2013). Future contemporaneous measurements of
radio photospheres and maser emission from various species and
transitions would be a powerful combination
for constraining the roles of magnetic fields and
their roles in shaping both the photosphere and the stellar outflow
over larger scales
(see also Gray et al. 2009; Amiri et al. 2012).

The data we have presented here have shown 
compelling evidence of a nonuniform, asymmetric brightness
distribution on the radio-emitting surface of Mira. Evidence for brightness
asymmetries on Mira has also been reported previously based on data at a variety
of other wavelengths (Haniff et al. 1995; Karovska 1999; Stewart et
al. 2015).  Lim et al. (1998) also showed that the red supergiant
Betelgeuse ($\alpha$~Ori) exhibits large-scale irregularities in its
radio surface (see also Richards
et al. 2013).
A likely cause of these brightness asymmetries are large-scale convective
cells, which are expected to be present on the surfaces of both AGB
stars and red supergiants (Schwarzschild 1975; Chiavassa \& Freytag
2015).

\section{Conclusions}
Using data obtained with the JVLA and ALMA, we 
have presented resolved measurements of the millimeter wavelength
continuum emission from both components of the Mira~AB binary system at
46~GHz (7~mm), 96~GHz (3~mm), and 229~GHz (1~mm). The millimeter
wavelength flux
densities measured for Mira~A are consistent with the radio
photosphere model previously derived by RM97. A comparison with
previously published measurements of Mira~A at 43~GHz
suggests that its flux density may vary over the course of the
stellar pulsation cycle and/or on timescales of years. However, additional
measurements are needed to confirm this possibility. The emission from
Mira~B is consistent to first order with free-free emission from a
circumstellar hypercompact \HII\
region of radius $\approx2\times10^{13}$~cm, although the millimeter spectrum
exhibits an unexpected upturn at higher frequencies, implying that a
more complex model is needed to fully describe the emission.

The stellar disk of Mira~A is resolved at all three observed
frequencies. We measure a mean photospheric radius of $\sim$3.0~AU at
46~GHz and find that the radius decreases systematically with
increasing frequency, as expected if the opacity decreases at shorter
wavelengths. The shape of the radio photosphere of Mira exhibits an
elongation, with a flattening of $\sim$10-20\%. It is presently
unclear whether the shape and/or the direction of the elongation evolves with time.

The data presented here reveal evidence of
brightness non-uniformities on the radio-emitting surface of
Mira~A. The cause of these non-uniformities may be large-scale
convective cells, similar to what has been previously proposed to explain the
surface irregularities on the
supergiant Betelgeuse. 

{\it Note added in proof:} After this paper was submitted, an
independent analysis of the ALMA data for Mira~AB by Vlemmings et
al. (2015) was submitted and accepted for
publication.   The analysis of the ALMA data by these authors differs from our own in that
for each ALMA band, Vlemmings et al. analyzed the data
from the individual observing dates separately, and they adopted
slightly different flux densities for the ALMA Band~3 flux
calibrators. They also performed
their imaging and data analysis using CASA and a stand-alone
$u$-$v$ fitting program rather than AIPS. 
The Vlemmings et al. study independently confirms evidence for a hot spot on
Mira~A and shows that the radio-emitting surface of Mira~B has now been
resolved. However, there are also some notable differences between our
findings. First, based on the results of
fits to the Band~3 visibility data, there are statistically
significant differences in the size and PA for Mira~A 
between our respective studies; indeed,
Vlemmings et al. derive a smaller size for Mira~A in Band~3 than in
Band~6, the opposite of what we report here. This in turn leads to a rather
different estimate for the Band~3 brightness temperature. While the respective sizes
that we derive for Mira~A in Band~6 agree to within uncertainties, the
PA of the major axis differs by more than 90$^{\circ}$. We note that
our value is consistent with the visible
elongation of the star in the image plane (e.g., Figure~3). 
Another important distinction between our results 
is that Vlemmings et al. report a shallower spectral
index for Mira~A than  we have advocated in the present work. They
also find evidence that the spectral index changes significantly
between 94 and 229~GHz. In contrast, we have found a single spectral
index of $\alpha$=1.86 (i.e., the value predicted by the radio photosphere model of MR97)
to be consistent with recent measurements  between 46 and  338~GHz.
Lastly, while both of our studies have revealed evidence for the presence
of a hot spot on Mira~A at approximately the same location, Vlemmings et al. report
a larger peak residual for this feature after subtracting the best
uniform disk fit and infer a smaller size for the spot. However, since
the inferred size of this feature is only a fraction of the
synthesized beam, its size, and hence its inferred
brightness temperature, are  necessarily
rather uncertain.


\acknowledgements
The authors thank T. Kami\'nski for valuable discussions and
E. Greisen for developments in AIPS that aided this work.
The JVLA observations presented here were part of NRAO program 14A-026.
This paper makes use of the following ALMA data:
ADS/JAO.ALMA \#2011.0.00014.SV. ALMA is a partnership of ESO 
(representing its member states), 
NSF (USA) and NINS (Japan), together with NRC (Canada) and NSC and 
ASIAA (Taiwan), in cooperation with the Republic of Chile. 
The Joint ALMA Observatory is operated by ESO, AUI/NRAO and NAOJ. 
The National Radio Astronomy Observatory is a facility of the National 
Science Foundation operated under cooperative agreement by Associated 
Universities, Inc.

\appendix
\section{Calculation of Uncertainties for Measured Quantities}
The error bars quoted for the measured quantities in Tables~2 and 3
contain contributions from both formal fitting uncertainties and systematic errors, where
the total error is assumed to be $\sigma_{t}=(\sigma^{2}_{f} + \sigma^{2}_{s})^{0.5}$. The
term $\sigma_{f}$ was taken to be the uncertainties reported by the AIPS {\small \sc{JMFIT}}
and {\small \sc{OMFIT}} tasks that were 
used for fitting
Gaussian and uniform disk models, respectively, and which
compute uncertainties according to the formalism outlined by Condon
(1997). 
The systematic errors,
$\sigma_{s}$, that affect our measurements include terms resulting
from blurring of the
images from
phase fluctuations on long baselines (i.e., ``seeing'' effects;
hereafter $\sigma_{s,p}$), as well as 
contributions that are difficult to quantify from first
principles (hereafter $\sigma_{s,o}$), such as those caused by the limited sampling of the
$u$-$v$ plane, by the size and shape of the interferometer dirty beam, 
and by the gridding and deconvolution process. 

Assuming that the residual phase noise on long baselines arises from 
tropospheric phase fluctuations that follow a Gaussian
distribution, then the true size of a source will be affected by convolution
with a Gaussian whose FWHM is given by $\theta_{p}=a\sqrt{8({\rm ln~2})}$,
where $a=\sigma_{\phi}\lambda/(2\pi b)$ (Thompson et al. 1994). Here
$\sigma_{\phi}$ represents the rms phase fluctuation on a baseline of
projected length $b$.  For the JVLA data ($\lambda\approx7$~mm) we find
$\sigma_{\phi}\approx29^{\circ}$ for the three longest baselines
($b$=34.5~km), implying $\theta_{p}\approx$7.5~mas. For the ALMA Band~3
data ($\lambda\approx$3~mm), $\sigma_{\phi}\approx75^{\circ}$ on a
baseline with $b$=10.66~km, implying
$\theta_{p}\approx29$~mas. Finally, for ALMA Band~6
($\lambda\approx$1.3~mm), $\sigma_{\phi}\approx96^{\circ}$ for
$b$=15.2~km, implying $\theta_{p}\approx$4.7~mas. Consequently, for our size
measurements we include in the error budget a relative error term
$\sigma_{s,p}
=[\theta_{m} - \sqrt{\theta^{2}_{m} - \theta_{p}^2}]/\theta_{m}$ where
$\theta_{m}$ is the measured major or minor axis FWHM. 
Phase noise also causes a decrease in the correlated amplitude on the
longest baselines of magnitude ${\rm exp}(-
\sigma_{\phi}^{2}/2)$ (e.g., Thompson et al.). 
However, the effect on the total integrated flux density of the
source in our case is quite small, and we have neglected this effect.

To obtain an
estimate of additional contributions $\sigma_{s,o}$ to the total systematic error budget, we 
performed a series of numerical experiments.
As an initial test,
we constructed a series of one-dimensional (1-D)
Gaussian signals, added to them different levels of random noise, and measured the apparent
FWHM. We then deconvolved the apparent width from the ``true'' width by subtracting in
quadrature a Gaussian representing various observing resolutions
to yield an intrinsic FWHM. Finally, based on series of 1000
trials for each combination of parameters (which bracketed the range
of signal-to-noise ratios and source-to-beam size ratios encountered
in our real data), we computed the rms
dispersion in the inferred signal width. In this way we have obtained a
lower limit to the expected systematic uncertainty in the measured source sizes.

In a second series of experiments, we introduced into our actual
visibility data at each frequency a   series 
of artificial elliptical Gaussians and uniform elliptical disks, offset by $-0.25$ arcseconds and
$-1.0$ arcseconds  south of
the phase center, respectively. For the Gaussian sources, we set the flux
density and axial ratio to the values measured from our fits to Mira~A (Table~2), but we
systematically varied the position angle from $-90^{\circ}$ to $+90^{\circ}$ degrees, in
increments of 15$^{\circ}$. For the uniform elliptical disks, we performed
trials using disk major and minor axes equal to 
our measured values from Table~2, but again varied the position
angles over several different values.
We then measured the parameters of
the artificial Gaussians in deconvolved
images using {\small {\sc JMFIT}} and of the artificial disks in the
visibility domain using {\small {\sc OMFIT}},
respectively, and compared these numbers with the
expected values. 

Not surprisingly,
we found the errors on the source sizes to be comparable to, or slightly higher in our 
2-D experiments compared with those derived from our initial 1-D experiments.
We also found that the 2-D Gaussian fits tended to systematically
underestimate the expected source flux density and size, typically by
a few per cent. This effect
persisted even if smaller cell sizes were used to compute the
deconvolved images. This effect may therefore arise
in part from the intrinsically non-Gaussian shape of the real
interferometer beam and/or deconvolution biases. 

Based on the above numerical experiments, we
adopt as $\sigma_{s,o}$ for
the image plane and visibility domain measurements, respectively, the mean
error in each measured quantity
from all of the 2-D trials at each of the three frequencies.
For
the sources sizes, the {\em total} systematic error includes contributions
both from $\sigma_{s,o}$ and $\sigma_{s,p}$ and is taken to be
$\sigma_{s}=(\sigma_{s,p}^{2} + \sigma_{s,o}^{2})^{0.5}$. 
In the case of the flux density,
we have additionally included contributions to the systematic error
budget from pointing uncertainties and absolute flux density
calibration (assumed to contribute relative errors of 1\% and 20\%,
respectively, at all three observed frequencies).

For Mira~A, the uncertainties on the source size and position
angle reported in this work are dominated by systematic uncertainties, whereas for the
weaker radio source Mira~B,
statistical errors dominate at two of the three observed frequencies. 
For both stars, the uncertainty in the integrated
flux density is dominated by the absolute calibration uncertainty
inherent at millimeter wavelengths.

\clearpage

%

%

%

%

%

%

%

\end{document}